\documentclass[12pt]{iopart}
\usepackage{iopams}

\begin{document}

\title[Phase-space descriptions of operators and the Wigner distribution I]{Phase-space descriptions of operators and the Wigner distribution in quantum mechanics I. \mbox{A Dirac} inspired view}

\author{S Chaturvedi\dag, E Ercolessi\ddag\footnote{To
whom correspondence should be addressed (ercolessi@bo.infn.it)}, G Marmo\S, G Morandi$\|$, \mbox{N Mukunda}\P\ and R Simon$^+$}

\address{\dag School of Physics, University of Hyderabad, Hyderabad 500 046, India}

\address{\ddag Physics Department, University of Bologna, INFM-CNR and INFN, Via Irnerio 46, I-40126, Bologna, Italy}

\address{\S Dipartimento di Scienze Fisiche, University of Napoli and INFN, Via Cinzia, I-80126 Napoli, Italy}

\address{$\|$ Physics Department, University of Bologna, INFM-CNR and INFN, V.le B.Pichat 6/2, I-40127, Bologna, Italy}

\address{\P Centre for High Energy Physics, Indian Institute of Science, Bangalore 560 012, India}

\address{$^+$ The Institute of Mathematical Sciences, C.I.T. Campus, Chennai 600113, India}

\begin{abstract}
Drawing inspiration from Dirac's work on functions of non commuting
observables, we develop a fresh approach to phase space descriptions of
operators and the Wigner distribution in quantum mechanics. The
construction presented here is marked by its economy,
naturalness and more importantly, by its potential for extensions and
generalisations to situations where the underlying configuration space
is non Cartesian.

\end{abstract}



\maketitle

\section{Introduction}

The development of classical phase space methods to describe quantum
systems whose kinematics is governed by Cartesian variables has its origin in
two important and independent ideas. The first, due to Weyl \cite{Wey} is the setting up of an unambiguous rule or convention that maps each (real)
classical dynamical variable, a classical phase-space function, into a
corresponding (hermitian) operator for the quantum system in a linear manner.
The second, due to Wigner \cite{Wig}, is the definition of a (real)
phase-space distribution function representing each (pure or mixed) quantum
state in a complete manner.  Later work \cite{Gro}  clarified that
these two rules or definitions are exactly inverses of one another, as a
result of which quantum mechanical expectation values can be written in a
classical-looking form as phase-space integrals.

In an interesting paper developing the analogies between classical and
quantum mechanics, Dirac \cite{Dir} discussed the general problem of
expressing a quantum dynamical variable, an operator, as a function of the
basic complete and irreducible set of operators of the quantum system. The
latter forms a non-commutative set, leading to the concept of ordering rules
in forming functions of them. In this context, Dirac used a description of an
operator by a collection of mixed matrix elements, the rows and columns
labelled by two different orthonormal bases for the Hilbert space of the system.

Phase space distributions have played a significant role in optics, particularly in unifying radiometry and radiative transfer with the theory of partial coherence. These activities were triggered by the pioneering work of Walther on radiometry and that of Wolf on radiative transfer. Walther introduced two definitions for the radiance function. The first definition \cite{Wal1} is analogous to the Wigner distribution and, indeed, the reader's attention was drawn by the author to this fact. Walther's second definition \cite{Wal2}, which has since been used in hundreds of radiometry papers may be seen, in retrospect, to be analogous to the Dirac inspired view of phase space distributions developed here. Interestingly, this remark applies equally well to Wolf's expression for the specific intensity \cite{Wol},  having been inspired by the second definition of Walther's.

The purpose of the present paper is to show that one can arrive at the
Weyl-Wigner formalism and results starting from Dirac's ideas and following
a physically well-motivated and at the same time an extremely elementary and
transparent series of steps. This illuminates the use of phase-space language
for quantum mechanics from an alternative perspective. The properties of
Wigner distributions have been studied in considerable detail by many authors.
It has been shown for instance, that if one asks for a phase-space description
of a quantum state obeying a small number of very reasonable conditions, then
the Wigner distribution is the unique answer. In important work by several
authors \cite{Cah} the Weyl-Wigner formalism has been shown to be
one of several different possibilities all based on the use of phase-space
methods. In the light of this, an alternative line of argument leading
directly and very easily to the same answer may be of interest.

Among the important properties of the Wigner distribution
corresponding to a given quantum state is that one recovers the two quantum
probability distributions in position or in momentum space by integrating over
the momentum or the position variable respectively, in the Wigner
distribution. This is called the recovery of the marginals. In our treatment
this too becomes immediately evident upon inspection and completely
transparent, no calculations being needed at all. The value of expressing
things in this way is that it may suggest possible generalizations to other
quantum-mechanical situations where the basic variables and commutation
relations are not of the Cartesian Heisenberg type. We have in mind for
instance the case of finite-dimensional  quantum systems \cite{Woo}, quantum mechanics on Lie groups \cite{MA}, non-commutative
geometric quantum schemes, deformed or $q$-quantum kinematics etc.  In all
these cases the presentation of the very well known and familiar case in a
very concise and transparent manner may help suggest interesting modifications
to be studied.

A brief summary of the present work is as follows. After a quick review
of the familiar phase space descriptions of operators in quantum
mechanics in Section 2,  in Section 3 we examine the question as to
how the trace of products of two operators can be expressed as
phase-space integrals in terms of their phase space representatives in
such a way that the inherent symmetry of the trace operation is
manifest. We show that this can be done at the expense of introducing a
kernel. We investigate the properties of this kernel in detail and show 
how by defining  a square root  of the kernel in a special way one is
naturally lead to the definition of Wigner distribution and associated
phase point operators. Some consequences of the Wigner-Weyl
correspondence are examined in Section 4, followed by a discussion on
the recovery of the marginal distributions in Section 5. Section 6
contains our concluding remarks and further outlook.

\section{Elementary Phase-Space Descriptions of Operators}

We consider a one-dimensional quantum system whose basic operators are
a hermitian Cartesian pair $\widehat{q},\widehat{p}$ obeying the Heisenberg
commutation relation:%
\begin{equation}
\left[  \widehat{q},\widehat{p}\right]  =i\hbar.
\end{equation}
For corresponding classical phase-space variables, as well as particular
classical values or quantum eigenvalues, we use $q,q^{\prime},q^{\prime\prime
},\cdots,p,p^{\prime},p^{\prime\prime},\cdots$. The continuum-normalized eigenstates
$\left\vert q\rangle,\left\vert p\rangle\; \right.  \right.  $ of
\ $\widehat{q}$ and $\widehat{p}$ obey as usual:%
\begin{equation}
\langle q|q^{\prime}\rangle =\delta\left(  q-q^{\prime}\right)
,\; \langle p|p^{\prime}\rangle =\delta\left(  p-p^{\prime
}\right)  ,\;\langle q|p\rangle =\left(  2\pi\hbar\right)
^{-1/2}\exp\left(  iqp/\hbar\right).  \label{norm}
\end{equation}
From the completeness relations:%
\begin{equation}%
{\displaystyle\int\limits_{-\infty}^{+\infty}}
dq|q\rangle\langle q|=%
{\displaystyle\int\limits_{-\infty}^{+\infty}}
dp|p\rangle\langle p|=\mathbb{I},
\end{equation}
we obtain immediately the two operator statements:%
\begin{equation}
\delta\left(  \widehat{q}-q\right)  =|q\rangle\langle q|,\;\delta\left(
\widehat{p}-p\right)  =|p\rangle\langle p|.
\end{equation}

Consider now a quantum-mechanical operator $\widehat{A}$. It can certainly
be completely described by its position-space matrix elements $\langle
q^{\prime}|\widehat{A}|q \rangle $ which constitute in general a
non-local kernel (In case $\widehat{A}$ is unitary, \ $\langle q^{\prime
}|\widehat{A}|q\rangle $ is the overlap between eigenstates of the ``old"
position operator \ $\widehat{q}$ and a ``new" one: \ $\widehat{q}^{\prime
}=\widehat{A}\widehat{q}\widehat{A}^{-1}$, and then the kernel of
\ $\widehat{A}$ would be the exponential of ($i$ times) the analogue of the
classical generating function of a canonical transformation of the type
``$q-Q$" \cite{Dir2}). The kernel corresponding to $\widehat{A}^{\dag
}$ is:%
\begin{equation}
\langle q^{\prime}|\widehat{A}^{\dag}|q\rangle =\langle
q|\widehat{A}|q^{\prime}\rangle ^{\ast}.%
\end{equation}

To move towards a description of $\widehat{A}$ at a classical phase-space
level it is natural to consider, in the spirit of Dirac \cite{Dir2},
the mixed matrix element \ $\langle q|\widehat{A}|p\rangle $ which,
regarded as a function of the phase-space variables $q$ and $p$, certainly
also describes $\widehat{A}$ completely. For later convenience we include a
non-vanishing plane wave factor and define the ``left" phase-space
representative of $\widehat{A}$ as the function:%
\begin{eqnarray}
A_{l}\left(  q,p\right)  &=& \langle q|\widehat{A}|p\rangle
\langle p|q\rangle =Tr\left\{  \widehat{A}|p\rangle\langle
p|q\rangle\langle q|\right\}  \nonumber \\
&=& Tr\left\{  \widehat{A}\delta\left(  \widehat{p}-p\right)  \delta\left(
\widehat{q}-q\right)  \right\} \label{Aleft}  \\
&=&\left(  2\pi\hbar\right)  ^{-1/2}\langle q|\widehat{A}|p\rangle \exp\left(  -iqp/\hbar\right). \nonumber
\end{eqnarray}
Here, of course, $\langle p|q\rangle $ is the kernel of the unitary
operator corresponding to Fourier transformation, which interchanges
$\widehat{q}$ and $\widehat{p}$. It is interesting to note that in Dirac's
treatment \cite{Dir2} $\ A_{l}\left(  q,p\right)  $ is regarded
essentially as the \textit{ratio} \ $\langle q|\widehat{A}|p\rangle
/\langle q|p\rangle $, which determines the form of $\widehat{A}$
as a function of \ $\widehat{q}$ and \ $\widehat{p}$ in standard ordered form,
i.e. $\widehat{q}$ to the left of $\widehat{p}$. \ 

Even if \ $\widehat{A}$ is hermitian, $A_{l}\left(  q,p\right)  $ is in
general complex. However we do have, as is particularly obvious from the
second line in Eq.(\ref{Aleft}):%
\begin{equation}
\eqalign{\int dpA_{l}\left(  q,p\right)  =\langle q|\widehat{A}|q\rangle
,\; \int dqA_{l}\left(  q,p\right)  =\langle p|\widehat
{A}|p\rangle , \\
Tr \left\{  \widehat{A}\right\}  =
\int \! \int
dqdpA_{l}\left(  q,p\right).}
\label{tr1}
\end{equation}

As an alternative to the above, the ``right" phase-space representative of
$\widehat{A}$ is given by:%
\begin{eqnarray}
A_{r}\left(  q,p\right)  &=&\langle p|\widehat{A}|q\rangle
\langle q|p\rangle =Tr\left\{  \widehat{A}|q\rangle\langle
q|p\rangle\langle p|\right\}  \nonumber \\
&=& Tr\left\{  \widehat{A}\delta\left(  \widehat{q}-q\right)  \delta\left(
\widehat{p}-p\right)  \right\}  \label{Aright} \\
&=& \left(  2\pi\hbar\right)  ^{-1/2}\langle p|\widehat{A}|q\rangle\exp\left(  iqp/\hbar\right).  \nonumber
\end{eqnarray}
This is related to expressing $\widehat{A}$ in anti-standard form,
\ $\widehat{p}$ to the left of \ $\widehat{q}$, and we have again:%
\begin{equation}
\eqalign{\int dpA_{r}\left(  q,p\right)  =\langle q|\widehat{A}|q\rangle
,\;\int dqA_{r}\left(  q,p\right)  =\langle p|\widehat
{A}|p\rangle, \\
Tr\left\{  \widehat{A}\right\}  =\int \! \int dqdpA_{r}\left(  q,p\right).}
\label{tr2}%
\end{equation}
Thus we have two equally elementary phase-space descriptions of the operator
$\widehat{A}$ on the same footing, with the r\^{o}les of coordinate and
momentum interchanged \ to go from one to the other. As noted above, even in
the hermitian case in general both $A_{l}\left(  q,p\right)  $ and
\ $A_{r}\left(  q,p\right)  $ are complex. More generally under hermitian
conjugation we have:%
\begin{equation}
\widehat{B}=:\widehat{A}^{\dag}\Rightarrow B_{r}\left(  q,p\right)
=A_{l}\left(  q,p\right)  ^{\ast},%
\end{equation}
so in the hermitian case we have:%
\begin{equation}
\widehat{A}^{\dag}=\widehat{A}\Rightarrow A_{r}\left(  q,p\right)
=A_{l}\left(  q,p\right)  ^{\ast}.%
\end{equation}

We can now ask if we can pass in a natural way to a third phase-space
description of \ $\widehat{A}$ standing exactly ``midway" \ between
$A_{l}\left(  q,p\right)  $ and $A_{r}\left(  q,p\right)  $, thus treating
$\widehat{q}$ and $\widehat{p}$ \ symmetrically. This is achieved in the next Section.

\section{Operator Product Traces and Passage to the Weyl-Wigner Description}

Consider two generally non-commuting operators $\widehat{A}$ and
$\widehat{B}$. The trace of their product is symmetric under their
interchange and can be expressed in two ways using classical phase
space\footnote{Thus, as is well known, from the point of view of trace
calculations the standard and anti-standard orderings are dual to one
another.}:%
\begin{eqnarray}
Tr\left\{  \widehat{A}\widehat{B}\right\}  &=& \int \! \int
dqdp\langle q|\widehat{A}|p\rangle \langle p|\widehat
{B}|q\rangle \nonumber \\
&=&2\pi\hbar \int \! \int
dqdpA_{l}\left(  q,p\right)  B_{r}\left(  q,p\right)  \\
&=& 2\pi\hbar \int \! \int
dqdpA_{r}\left(  q,p\right)  B_{l}\left(  q,p\right). \nonumber 
\end{eqnarray}
The last line follows from the previous one by symmetry under interchange of
$\widehat{A}$ and $\widehat{B}$. However in each of these two phase-space
integrals the manifest symmetry in \ $\widehat{A}$ and $\widehat{B}$ is
lacking. One can ask if such \ symmetry can be restored while continuing to
work with phase-space quantities. \ Towards this end we begin by first
expressing \ $(\widehat{A}\widehat{B})_{l}\left(  q,p\right)  $ entirely in
terms of $A_{l}\left(  q^{\prime},p^{\prime}\right)  $ and $B_{l}\left(
q^{\prime\prime},p^{\prime\prime}\right)  $:%
\begin{eqnarray}
\fl \left(  \widehat{A}\widehat{B}\right)  _{l}\left(  q,p\right)  &=&\langle
q|\widehat{A}\widehat{B}|p\rangle \langle p|q\rangle 
= \int \! \int 
dq^{\prime}dp^{\prime}\langle q|\widehat{A}|p^{\prime}\rangle
\langle p^{\prime}|q^{\prime}\rangle \langle q^{\prime
}|\widehat{B}|p\rangle \langle p|q\rangle \nonumber\\
&=& \int \! \int 
dq^{\prime}dp^{\prime}A_{l}\left(  q,p^{\prime}\right)  K_{l}\left(
q,p^{\prime};q^{\prime},p\right)  B_{l}\left(  q^{\prime},p\right) \label{eq3.2}, \\
\fl K_{l}\left(  q,p^{\prime};q^{\prime},p\right)  &=&\left(  2\pi\hbar\right)
^{2}\langle q|p^{\prime}\rangle \langle p^{\prime}|q^{\prime
}\rangle \langle q^{\prime}|p\rangle \langle
p|q\rangle 
=\exp\left\{  i\left(  q-q^{\prime}\right)  \left(  p^{\prime}-p\right)
/\hbar\right\}. \nonumber
\end{eqnarray}
the first line in the definition of $K_{l}$ \ following from: $\langle
q|\widehat{A}|p^{\prime}\rangle =A_{l}\left(  q,p^{\prime}\right)
/\langle p^{\prime}|q\rangle $ and from: $1/\langle p^{\prime
}|q\rangle =2\pi\hbar\langle q|p^{\prime}\rangle $.
The non-local convolution involved in expressing $(\widehat{A}\widehat{B}%
)_{l}$ in terms of $A_{l}$ and $B_{l}$ \ is an indication already of the
general situation since we are dealing with mixed matrix elements; it is a
forerunner \ of the Moyal or ``star" product when the transition to the
Weyl-Wigner description of operators is completed. We may also note that,
aside from the continuum normalization of the $\widehat{q}$ \ and
\ $\widehat{p}$ eigenvectors, \ the kernel $K_{l}$ is a four-vertex Bargmann
invariant \cite{Bar}. \ Hence its phase, which is the area \ of the
phase-space rectangle with vertices $\left(  q,p\right)  ,\left(  q,p^{\prime
}\right)  ,\left(  q^{\prime},p\right)  $ and $\left(  q^{\prime},p^{\prime
}\right)  $ is a geometric phase \cite{Ber}. \ Now combining
Eq.(\ref{eq3.2}) with Eq.(\ref{tr1}) and relabelling some variables for
convenience, we get \ $Tr\{\widehat{A}\widehat{B}\}$ entirely in terms of left
representatives:%
\begin{equation}
Tr\left\{  \widehat{A}\widehat{B}\right\}  = \int \! \int \! \int \! \int 
dqdpdq^{\prime}dp^{\prime}A_{l}\left(  q,p\right)  K_{l}\left(  q,p;q^{\prime
},p^{\prime}\right)  B_{l}\left(  q^{\prime},p^{\prime}\right).  \label{trace}%
\end{equation}
The kernel \ $K_{l}\left(  q,p;q^{\prime},p^{\prime}\right)  $ is explicitly
symmetric under: $\left(  q,p\right)  \longleftrightarrow\left(  q^{\prime
},p^{\prime}\right)  $, so we have a classical phase-space expression for
\ $Tr\{\widehat{A}\widehat{B}\}$ manifestly symmetric in $\widehat{A}$ and
\ $\widehat{B}$, but at the cost of a kernel. In addition to symmetry, this
kernel possesses two important properties: it is invariant under phase-space
translations as it depends only on the differences $q-q^{\prime}$,
$p-p^{\prime}$; and it satisfies the ``marginals" equations:%
\begin{equation}
\eqalign{
\int dp^{\prime}K_{l}\left(  q,p;q^{\prime},p^{\prime}\right)  =2\pi
\hbar\delta\left(  q-q^{\prime}\right), \\
\int dq^{\prime}K_{l}\left(  q,p;q^{\prime},p^{\prime}\right)  =2\pi
\hbar\delta\left(  p-p^{\prime}\right).}
\label{alpha}%
\end{equation}
The most natural question is to ask if this kernel can in some sense be
``transformed away" while maintaining manifest symmetry in  $\widehat{A}$ and $\widehat{B}$. This can be done if we can express it as the ``square" or the convolution of some more elementary kernel, say in the form:%
\begin{equation}
K_{l}\left(  q,p;q^{\prime},p^{\prime}\right)  = \int \! \int 
dq^{\prime\prime}dp^{\prime\prime}\xi\left(  q^{\prime\prime},p^{\prime\prime
};q,p\right)  \xi\left(  q^{\prime\prime},p^{\prime\prime};q^{\prime
},p^{\prime}\right).  \label{beta}%
\end{equation}
From the known properties of \ $K_{l}\left(  q,p;q^{\prime},p^{\prime}\right)
$ we can demand that $\xi\left(  q,p;q^{\prime},p^{\prime}\right)  $ too be a
symmetric function of its (pairs of) arguments; be invariant under phase-space
translations and so depend only on the differences $q-q^{\prime}$,
$p-p^{\prime}$; and possess the ``marginals" property:%
\begin{equation}%
\eqalign{\int dp^{\prime}\xi\left(  q,p;q^{\prime},p^{\prime}\right)  =\sqrt{2\pi\hbar
}\delta\left(  q-q^{\prime}\right), \\
\int dq^{\prime}\xi\left(  q,p;q^{\prime},p^{\prime}\right)  =\sqrt{2\pi\hbar
}\delta\left(  p-p^{\prime}\right).}
\label{gamma}%
\end{equation}
Easy calculation shows that the expression:%
\begin{equation}
\xi\left(  q,p;q^{\prime},p^{\prime}\right)  =\sqrt{\frac{2}{\pi\hbar}}%
\exp\left\{  2i\left(  q-q^{\prime}\right)  \left(  p-p^{\prime}\right)
/\hbar\right\}  \label{xi}%
\end{equation}
obeys all the conditions imposed above\footnote{It is interesting that while
as a function of four phase-space variables $\xi$ is the pointwise
\textit{square} of $K_{l}$, as a kernel it is the \textit{square root} of
$K_{l}$.}.
If we use this in Eq.(\ref{trace}) and associate one factor of $\xi$ each with
$A_{l}$ and $B_{l}$ we arrive at the simpler expression:%
\begin{equation}
Tr\left\{  \widehat{A}\widehat{B}\right\}  =\frac{1}{2\pi\hbar} \int \! \int
dqdpA\left(  q,p\right)  B\left(  q,p\right),  \label{trace2}%
\end{equation}
where \ $A\left(  q,p\right)  $ arises from $A_{l}\left(  q,p\right)  $ via:%
\begin{equation}
A\left(  q,p\right)  =\sqrt{2\pi\hbar} \int \! \int 
dq^{\prime}dp^{\prime}\xi\left(  q,p;q^{\prime},p^{\prime}\right)
A_{l}\left(  q^{\prime},p^{\prime}\right)  \label{eqA}%
\end{equation}
and similarly for $B\left(  q,p\right)  $\footnote{The factor in
Eq.(\ref{trace2}) is chosen so that \ $\widehat{A}$ and $A\left(  q,p\right)$ have the same physical dimension.}.  We have thus achieved, by a two-step
procedure, our objective of expressing $Tr\{\widehat{A}\widehat{B}\}$ as a
manifestly symmetric classical phase-space integral, with one phase-space
function each representing $\widehat{A}$ and  $\widehat{B}$  and with no
additional kernel. One can now see that in the case  $\widehat{A}^{\dag
}=\widehat{A},\widehat{B}^{\dag}=\widehat{B}$, since  $Tr\{\widehat
{A}\widehat{B}\}$ is real and \ $\widehat{A}$  and $\widehat{B}$ can be
 chosen independently, $A\left(  q,p\right)  $ and $B\left(  q,p\right)  $
must be individually real:%
\begin{equation}
\widehat{A}=\widehat{A}^{\dag}\Rightarrow A\left(  q,p\right)  =A\left(
q,p\right)  ^{\ast}.%
\end{equation}

Expression (\ref{eqA}) for $A\left(  q,p\right)  $ is indeed the Weyl-Wigner
representative of \ $\widehat{A}$ \ in phase-space form. With elementary
manipulations we can express it \ in the equivalent forms \cite{FN1}:
\begin{eqnarray}
A\left(  q,p\right)  &=& 2 \int \! \int 
dq^{\prime}dp^{\prime}A_{l}\left(  q^{\prime},p^{\prime}\right)  \exp\left\{
2i\left(  q-q^{\prime}\right)  \left(  p-p^{\prime}\right)  /\hbar\right\}
\\
&=& \int \! \int 
dq^{\prime}\langle q-\frac{1}{2}q^{\prime}|\widehat{A}|q+\frac{1}%
{2}q^{\prime}\rangle \exp\left\{  ipq^{\prime}/\hbar\right\}
\end{eqnarray}
or:%
\begin{equation}
A\left(  q,p\right)  =2\pi\hbar Tr\left\{  \widehat{A}\widehat{\mathcal{W}%
}\left(  q,p\right)  \right\},  \label{Wigner1}%
\end{equation}
where:%
\begin{eqnarray}
\widehat{\mathcal{W}}\left(  q,p\right)  &=& \frac{1}{2\pi\hbar} \int \! \int 
dq^{\prime}|q+\frac{1}{2}q^{\prime}\rangle\langle q-\frac{1}{2}q^{\prime}
|\exp\left\{  ipq^{\prime}/\hbar\right\} \label{dabliu} \\
&=& \int \! \int 
dq^{\prime}|q+\frac{1}{2}q^{\prime}\rangle\langle q+\frac{1}{2}q^{\prime
}|p\rangle\langle p|q-\frac{1}{2}q^{\prime}\rangle\langle q-\frac{1}%
{2}q^{\prime}|. \nonumber 
\end{eqnarray}
Representing this ``phase-point operator" as:%
\begin{equation}
\widehat{\mathcal{W}}\left(  q,p\right)  =\int dq^{\prime}dq^{\prime\prime
}|q^{\prime}\rangle\langle q^{\prime}|\widehat{\mathcal{W}}\left(
q,p\right)  |q^{\prime\prime}\rangle \langle q^{\prime\prime}|,
\end{equation}
the matrix elements are given by:%
\begin{equation}
\langle q^{\prime}|\widehat{\mathcal{W}}\left(  q,p\right)
|q^{\prime\prime}\rangle =\frac{1}{\pi\hbar}\delta\left(  q^{\prime
}+q^{\prime\prime}-2q\right)  \exp\left\{  ip\left(  q^{\prime}-q^{\prime
\prime}\right)  /\hbar\right\}.  \label{matel}%
\end{equation}
It is evident from\ the first line in Eq.(\ref{dabliu}) that $\widehat
{\mathcal{W}}\left(  q,p\right)  $ is hermitian as well as that:%
\begin{equation}
Tr\left\{  \widehat{\mathcal{W}}\left(  q,p\right)  \right\}  =\frac{1}%
{2\pi\hbar}. \label{trace1}%
\end{equation}

The Weyl correspondence makes use (see below) of the ``Weyl operators"
$\exp\left\{  i\left(  x\widehat{p}-k\widehat{q}\right)  /\hbar\right\}  $
that are labelled by the phase-space points $x$ and $k$. Then, one can prove
that:%
\begin{equation}
\widehat{\mathcal{W}}\left(  q,p\right)  = \int \! \int 
\frac{dxdk}{(2\pi\hbar)^{2}}\exp\left\{  -i\left(  xp-kq\right)  \right\}
\exp\left\{  i\left(  x\widehat{p}-k\widehat{q}\right)  /\hbar\right\},
\label{fourier}%
\end{equation}
i.e. that the phase-point operators are the symplectic Fourier transforms of
the Weyl operators. Indeed, a straightforward calculation shows that:%
\begin{equation}
\langle q^{\prime}|\exp\left\{  i\left(  x\widehat{p}-k\widehat
{q}\right)  /\hbar\right\}  |q^{\prime\prime}\rangle =\delta\left(
x+q^{\prime}-q^{\prime\prime}\right)  \exp\left\{  -ik\frac{q^{\prime
}+q^{\prime\prime}}{2\hbar}\right\}
\end{equation}
and using this result to evaluate the matrix elements of both sides of
Eq.(\ref{fourier}) one obtains back Eq.(\ref{matel}).

We look at some more properties of the operators \ $\widehat{\mathcal{W}%
}\left(  q,p\right)  $ \ in the next Section.

\section{Some Consequences of the Weyl-Wigner Correspondence}

It is illuminating to compute the phase-space \ functions representing the
operators $\delta\left(  \widehat{q}-q\right)  \delta\left(  \widehat
{p}-p\right)  ,\widehat{\mathcal{W}}\left(  q,p\right)  $ and \ $\delta\left(
\widehat{p}-p\right)  \delta\left(  \widehat{q}-q\right)  $ \ occurring in
Eqs.(\ref{Aleft}), (\ref{Aright}) and (\ref{dabliu}) under the Weyl-Wigner
correspondence. This helps us also understand the structure of the kernel
$\xi\left(  q,p;q^{\prime}p^{\prime}\right)  $ better. It is well known that
the Weyl correspondence takes elementary classical exponentials to elementary
operator exponentials according to:%
\begin{equation}
\exp\left\{  i\left(  \sigma q-\tau p\right)  \right\}  \rightarrow
\exp\left\{  i\left(  \sigma\widehat{q}-\tau\widehat{p}\right)  \right\}
\end{equation}
and then extends this to general functions $A\left(  q,p\right)  $ by
linearity. For the operators mentioned above we find using Eqs.(\ref{Aleft})
and (\ref{eqA}):%
\begin{eqnarray}
\fl \widehat{A}=\delta\left(  \widehat{q}-q^{\prime}\right)  \delta\left(
\widehat{p}-p^{\prime}\right)  &\Rightarrow& A_{l}\left(  q,p\right)  =\frac{1}{2\pi\hbar}\delta\left(  q-q^{\prime}\right)  \delta\left(  p-p^{\prime
}\right) \nonumber \\
&\Rightarrow& A\left(  q,p\right)  =\frac{1}{\pi\hbar}\exp\left\{  2i\left(
q-q^{\prime}\right)  \left(  p-p^{\prime}\right)  /\hbar\right\},\\
\fl \widehat{A}=\delta\left(  \widehat{p}-p^{\prime}\right)  \delta\left(
\widehat{q}-q^{\prime}\right)  &\Rightarrow& A\left(  q,p\right)  =\frac{1}%
{\pi\hbar}\exp\left\{  -2i\left(  q-q^{\prime}\right)  \left(  p-p^{\prime
}\right)  /\hbar\right\},
\end{eqnarray}
the  second result following from the first by hermitian conjugation, and we see that, up to a factor, the latter is just the kernel $\xi\left(
q,p;q^{\prime},p^{\prime}\right)  $ of Eq.(\ref{xi}). For $\widehat
{\mathcal{W}}\left(  q^{\prime},p^{\prime}\right)  $ we find from
Eq.(\ref{dabliu}):%
\begin{eqnarray}
\fl \widehat{A}=\widehat{\mathcal{W}}\left(  q^{\prime},p^{\prime}\right)
&\Rightarrow& A_{l}\left(  q,p\right)  =\frac{2}{\left(  2\pi\hbar\right)  ^{2} }\exp\left\{  -2i\left(  q-q^{\prime}\right)  \left(  p-p^{\prime}\right)
/\hbar\right\}  \nonumber \\
&\Rightarrow& A\left(  q,p\right)  =\delta\left(  q-q^{\prime}\right)
\delta\left(  p-p^{\prime}\right). \label{dabliu2}
\end{eqnarray}
At first sight, it is not easy to recognize that the operators $\widehat
{\mathcal{W}}\left(  q^{\prime},p^{\prime}\right)  $ stand ``midway" \ between
$\delta\left(  \widehat{q}-q^{\prime}\right)  \delta\left(  \widehat
{p}-p^{\prime}\right)  $ and \ $\delta\left(  \widehat{p}-p^{\prime}\right)
\delta\left(  \widehat{q}-q^{\prime}\right)  $, in the sense of treating
\ $\widehat{q}$ and $\widehat{p}$ symmetrically, or that the classical
representative \ $\delta\left(  q-q^{\prime}\right)  \delta\left(
p-p^{\prime}\right)  $ of \ \ $\widehat{\mathcal{W}}\left(  q^{\prime
},p^{\prime}\right)  $ also stands ``midway" as a phase-space function between
the two functions \ $\left(  1/\pi\hbar\right)  \exp\left\{  \pm2i\left(
q-q^{\prime}\right)  \left(  p-p^{\prime}\right)  /\hbar\right\}  $. However
this is actually so, as can be appreciated by looking at the Fourier
transforms with respect to $q^{\prime}$ and $p^{\prime}$. Classically we have:%
\begin{eqnarray}
\fl  \frac{1}{\left(  2\pi\right)  ^{2}} \int \! \int
d\sigma d\tau && \exp\left\{  i\left[  \sigma\left(  q-q^{\prime}\right)
-\tau\left(  p-p^{\prime}\right)  \right]  \right\}  \times\left\{
\exp\left(  \pm i\hbar\sigma\tau/2\right)  \; {\rm or} \;1\right\}  \nonumber \\
&&=\frac{1}{\pi\hbar}\exp\left\{  \pm2i\left(  q-q^{\prime}\right)  \left(
p-p^{\prime}\right)  /\hbar\right\}  \;{\rm or} \; \delta\left(
q-q^{\prime}\right)  \delta\left(  p-p^{\prime}\right)
\end{eqnarray}
and the Weyl map then preserves these ``relative positions" among the
corresponding operators.

Two other known properties of the $\widehat{\mathcal{W}}\left(  q,p\right)  $
are immediately read off from \ the relations assembled above, with no need
for any calculations. Since $\delta\left(  q-q^{\prime}\right)  \delta\left(
p-p^{\prime}\right)  $ is real, \ $\widehat{\mathcal{W}}\left(  q^{\prime
},p^{\prime}\right)  $ is hermitian, as we have already noted,and from
Eqs.(\ref{trace2}) and (\ref{dabliu2}) they are trace orthonormal in the
continuum sense:%
\begin{eqnarray}
\fl Tr\left\{  \widehat{\mathcal{W}}\left(  q,p\right)  \widehat{\mathcal{W}}\left(  q^{\prime},p^{\prime}\right)  \right\} & =&
\frac{1}{2\pi\hbar} \int \! \int 
dq^{\prime\prime}dp^{\prime\prime}\delta\left(  q-q^{\prime\prime}\right)
\delta\left(  p-p^{\prime\prime}\right)  \delta\left(  q^{\prime}%
-q^{\prime\prime}\right)  \delta\left(  p^{\prime}-p^{\prime\prime}\right)
\nonumber \\
&=&\frac{1}{2\pi\hbar}\delta\left(  q-q^{\prime}\right)  \delta\left(
p-p^{\prime}\right).
\label{trace3}%
\end{eqnarray}
Thus the inverse of Eq.(\ref{Wigner1}) reads:%
\begin{equation}
\widehat{A}= \int \! \int 
dqdpA\left(  q,p\right)  \widehat{\mathcal{W}}\left(  q,p\right).
\end{equation}

It is known \cite{Muk} that the operators $\widehat{\mathcal{W}%
}\left(  q,p\right)  $ obey the following interesting anti-commutation
relations with $\widehat{q}$ and  $\widehat{p}$:%
\begin{equation}
\frac{1}{2}\left\{  \widehat{q},\widehat{\mathcal{W}}\left(  q,p\right)
\right\}  =q\widehat{\mathcal{W}}\left(  q,p\right)  ,\; \frac{1}%
{2}\left\{  \widehat{p},\widehat{\mathcal{W}}\left(  q,p\right)  \right\}
=p\widehat{\mathcal{W}}\left(  q,p\right).  \label{anticom}%
\end{equation}
These are operator versions of corresponding immediately obvious classical
relations stemming from: $q\delta\left(  q-q^{\prime}\right)  =q^{\prime
}\delta\left(  q-q^{\prime}\right)  $. If we set $q=p=0$ in Eq. (\ref{anticom}%
) we see that $\widehat{\mathcal{W}}\left(  0,0\right)  $ anticommutes with
both $\widehat{q}$ and $\widehat{p}$:%
\begin{equation}
\widehat{q} \, \widehat{\mathcal{W}}\left(  0,0\right)  =-\widehat{\mathcal{W}%
}\left(  0,0\right)  \widehat{q},\; \widehat{p}\; \widehat{\mathcal{W}%
}\left(  0,0\right)  =-\widehat{\mathcal{W}}\left(  0,0\right)  \widehat{p}.
\label{anticom2}%
\end{equation}
In turn this means that  $\widehat{\mathcal{W}}\left(  0,0\right)  ^{2}$
commutes with both $\widehat{q}$ and  $\widehat{p}$, so it must be a multiple
of the unit operator. We can easily convince ourselves that $\widehat
{\mathcal{W}}\left(  0,0\right)  $ is nonzero, so we must be dealing here with
a nonzero multiple, which means that  $\widehat{\mathcal{W}}\left(
0,0\right)  $ has an inverse. The existence of the inverse together with the
first of Eqs.(\ref{anticom2}) leads at once to: $\widehat{\mathcal{W}}\left(
0,0\right)  \widehat{q}\, \widehat{\mathcal{W}}\left(  0,0\right)  ^{-1}%
=-\widehat{q}$ and similarly for $\widehat{p}$, so the upshot \ is that
$\widehat{\mathcal{W}}\left(  0,0\right)  $ is a multiple of the parity
operator. Now, e.g. in the $\left\{  |q\rangle\right\}  $ basis, the matrix
elements of the parity operator $\widehat{P}$ are given by (cfr.Eq.(\ref{norm}%
)):%
\begin{equation}
\langle q|\widehat{P}|q^{\prime}\rangle =\langle q|-q^{\prime
}\rangle =\delta\left(  q+q^{\prime}\right)
\end{equation}
and then, in this basis:%
\begin{equation}
Tr\left\{  \widehat{P}\right\}  =\int dq\langle q|\widehat{P}%
|q\rangle =\int dq\delta\left(  2q\right)  =\frac{1}{2}.%
\end{equation}
Taking then traces and using Eq.(\ref{trace1}) fixes the proportionality
factor and we find:%
\begin{equation}
\widehat{\mathcal{W}}\left(  0,0\right)  =\frac{1}{\pi\hbar}\widehat{P}.%
\end{equation}
Summarizing, one finds the Weyl correspondence:%
\begin{equation}
A\left(  q,p\right)  =\delta\left(  q\right)  \delta\left(  p\right)
\Rightarrow\widehat{A}=\widehat{\mathcal{W}}\left(  0,0\right)  =\frac{1}%
{\pi\hbar}\widehat{P}. \label{parity}%
\end{equation}

\textbf{Remark.} We would like to stress that what we mean here by ``trace"
of an operator is defined as the sum (or integral) of the diagonal matrix
elements of the operator with respect to a \textit{given} basis, and that the
\textit{only} bases that we are using here are the $\left\{  |q\rangle
\right\}  $ and $\left\{  |p\rangle\right\}  $ bases, in each of which, e.g.
and consistently: $Tr\{\widehat{P}\}=1/2$. This does \textit{not} mean that we
are claiming that $\widehat{P}$ is a trace-class operator, which would imply
much stronger requirements that the parity operator is not likely to meet.

One can carry the previous analysis one step further. The unitary operator:%
\begin{equation}
\widehat{U}\left(  q,p\right)  =:\exp\left\{  i\left[  p\widehat{q}%
-q\widehat{p}\right]  /\hbar\right\}
\end{equation}
acts as a phase-space displacement operator. Indeed:%
\begin{equation}
\widehat{U}\left(  q,p\right)  \widehat{q}\, \widehat{U}\left(  q,p\right)
^{\dag}=\widehat{q}-q,\; \widehat{U}\left(  q,p\right)  \widehat
{p}\, \widehat{U}\left(  q,p\right)  ^{\dag}=\widehat{p}-p
\end{equation}
and hence, for any (analytic) operator $\widehat{O}=\widehat{O}\left(
\widehat{q},\widehat{p}\right)  $:%
\begin{equation}
\widehat{U}\left(  q,p\right)  \widehat{O}\left(  \widehat{q},\widehat
{p}\right)  \widehat{U}\left(  q,p\right)  ^{\dag}=\widehat{O}\left(
\widehat{q}-q,\widehat{p}-p\right).
\end{equation}
Taking then: $\widehat{O}\left(  \widehat{q},\widehat{p}\right)
=\widehat{\mathcal{W}}\left(  0,0\right)  $ from Eq.(\ref{fourier}) one
obtains immediately:%
\begin{equation}
\widehat{U}\left(  q,p\right)  \widehat{\mathcal{W}}\left(  0,0\right)
\widehat{U}\left(  q,p\right)  ^{\dag}=\widehat{\mathcal{W}}\left(
q,p\right)
\end{equation}
and hence, from Eq.(\ref{parity}):%
\begin{equation}
\widehat{\mathcal{W}}\left(  q,p\right)  =\frac{1}{\pi\hbar}\widehat{U}\left(
q,p\right)  \widehat{P}\widehat{U}\left(  q,p\right)  ^{\dag}.%
\end{equation}
Thus the phase-point operators are just the parity operation with respect to general phase-space points. This leads to the at first sight unexpected
operator property:%
\begin{equation}
\widehat{\mathcal{W}}\left(  q,p\right)  ^{2}=\frac{1}{\left(  \pi
\hbar\right)  ^{2}}\mathbb{I}, \label{epsilon}%
\end{equation}
which means that the eigenvalues of \ $\widehat{\mathcal{W}}\left(
q,p\right)  $ are $\pm1/\pi\hbar$. This will be used in the next Section.

The purpose of this discussion was to show that this otherwise rather
unexpected fact is an immediate consequence of the relevant operator relations given above.

\section{Recovery of Marginal Distributions}

We have mentioned in Section 1 that an important property of the Wigner
distribution is that upon partial integration over either $p$ or $q$ the
complementary quantum-mechanical probability distribution emerges. In this
Section we show how this happens practically automatically or manifestly if we
use the relations given in Section 3.

The general Weyl association is as given in Eqs.(\ref{eqA}), (\ref{dabliu}).
For the density operator $\widehat{\rho}$ representing a (pure or mixed)
quantum state, we use a different normalization and define the Wigner
distribution by:%
\begin{equation}
\rho\left(  q,p\right)  =Tr\left\{  \widehat{\rho}\widehat{\mathcal{W}}\left(
q,p\right)  \right\},
\end{equation}
so that the general expression for the expectation value of $\widehat{A}$ in
state $\widehat{\rho}$ has the form:%
\begin{equation}
Tr\left\{  \widehat{\rho}\widehat{A}\right\}  = \int \! \int 
dqdpA\left(  q,p\right)  \rho\left(  q,p\right).
\end{equation}

We notice at this point that Eq.(\ref{epsilon}) concerning the spectrum of
\ $\widehat{\mathcal{W}}\left(  q,p\right)  $ has the following implication:%
\begin{equation}
\left\vert \rho\left(  q,p\right)  \right\vert \leq\frac{1}{\pi\hbar}.%
\end{equation}

This known property of Wigner distributions is usually obtained by using the
Cauchy-Schwartz inequality, so it is interesting to see it emerging here by a
much more elementary argument. Now we turn to the marginals.

For a general operator \ $\widehat{A}$ \ we have seen in Eqs.(\ref{tr1}) and
(\ref{tr2}) that:%
\begin{equation}
\int dpA_{l,r}\left(  q,p\right)  =\langle q|\widehat{A}|q\rangle
,\; \int dqA_{l,r}\left(  q,p\right)  =\langle p|\widehat
{A}|p\rangle.
\end{equation}
At the next step, for the kernel $\xi\left(  q,p;q^{\prime},p^{\prime}\right)
$ of Eq.(\ref{beta}) we have the properties in Eq.(\ref{gamma}).
Combining the above two pairs of equations with the passage (\ref{eqA}) from
$A_{l}\left(  q,p\right)  $ to the \ Weyl representative $A\left(  q,p\right)
$ of \ $\widehat{A}$, it is immediately seen that:%
\begin{equation}
\int dpA\left(  q,p\right)  =2\pi\hbar\langle q|\widehat{A}%
|q\rangle ,\; \int dqA\left(  q,p\right)  =2\pi\hbar\langle
p|\widehat{A}|p\rangle.
\end{equation}
Essentially no work has to be done to get these results. In \ the case of the
density operator $\widehat{\rho}$ we omit the factor $2\pi\hbar$, so we
recover the marginal probability distributions as:%
\begin{equation}
\int dp\rho\left(  q,p\right)  =\langle q|\widehat{\rho}|q\rangle
,\; \int dq\rho\left(  q,p\right)  =\langle p|\widehat{\rho
}|p\rangle.
\end{equation}

We can also obtain operator forms of such statements for \ $\widehat
{\mathcal{W}}\left(  q,p\right)  $. Knowing that the Weyl map is linear,
\ that $\widehat{\mathcal{W}}\left(  q^{\prime},p^{\prime}\right)  $
corresponds to $\delta\left(  q-q^{\prime}\right)  \delta\left(  p-p^{\prime}\right)  $, and that the operators corresponding to $\delta\left(
q-q^{\prime}\right)  $ and \ $\delta\left(  p-p^{\prime}\right)  $ are
$\delta\left(  \widehat{q}-q^{\prime}\right)  $ and \ $\delta\left(
\widehat{p}-p^{\prime}\right)  $, we have:%
\begin{eqnarray}
\fl \int dp^{\prime}\delta\left(  q-q^{\prime}\right)  \delta\left(  p-p^{\prime
}\right)  =\delta\left(  q-q^{\prime}\right)  &\Leftrightarrow&
\int dp^{\prime}\widehat{\mathcal{W}}\left(  q^{\prime
},p^{\prime}\right)  =\delta\left(  \widehat{q}-q^{\prime}\right)
=|q^{\prime}\rangle\langle q^{\prime}| ,\\
\fl \int dq^{\prime}\delta\left(  q-q^{\prime}\right)  \delta\left(  p-p^{\prime}\right)  =\delta\left(  p-p^{\prime}\right)  &\Leftrightarrow&
\int dq^{\prime}\widehat{\mathcal{W}}\left(  q^{\prime
},p^{\prime}\right)  =\delta\left(  \widehat{p}-p^{\prime}\right)
=|p^{\prime}\rangle\langle p^{\prime}|.
\end{eqnarray}
The purpose of showing the recovery of marginals in this manner is again to
emphasize how elementary and transparent the derivations are.

\section{Concluding Remarks}

To conclude, we have shown how by expressing the trace of
product of two operators in terms of their phase space representatives
and some fairly elementary steps one is  naturally led to the concept of
the Wigner distribution. Crucial to this construction is a kernel with
the structure of a Bargmann invariant and its square root endowed with
certain desirable properties. A noteworthy feature of the approach
developed here is its economy - no auxiliary constructs are required at
all and above all the facility with which it lends itself to
application to non Cartesian situations such as finite state quantum
systems which are of particular relevance to quantum computation and
quantum information  processing. That  this is indeed the case will be
demonstrated in a companion paper.


\section*{References}

\begin{thebibliography}{99}

\bibitem{Wey} Weyl H 1927 {\it Z. Phys.} \textbf{46} 1  and 1931  {\it The Theory of Groups and Quantum Mechanics} (Dover, N.Y.) p~274

\bibitem{Wig} Wigner E P 1932 {\it Phys. Rev.} \textbf{40} 749.  For
reviews see: Hillery M, O'Connell R F, Scully M O and Wigner E P 1984 {\it Phys. Repts.} \textbf{106} 121; Kim Y S and Noz M E 1991{\it Phase-Space Picture of Quantum Mechanics} (World Scientific, Singapore); Schleich W P 2001 {\it Quantum Optics in Phase Space} (Wiley-VCH, Weinheim)

\bibitem{Gro} Groenewold H J 1946 {\it Physica} \textbf{12} 205;
Moyal J E 1949 {\it Proc. Camb. Phil. Soc.} \textbf{45} 99 

\bibitem{Dir} Dirac P A M 1945 {\it Revs. Mod. Phys.} \textbf{17} 195 

\bibitem{Wal1} Walther A 1968 {\it  J. Opt. Soc. Am.}  \textbf{58} 1256

\bibitem{Wal2} Walther A 1973 {\it  J. Opt. Soc. Am.}  \textbf{63} 1622

\bibitem{Wol} Wolf E 1976 {\it  Phys. Rev. D} \textbf{13} 869

\bibitem{Cah} See, for instance, in addition to Refs. \cite{Wig}: Cahill K E and Glauber R J 1969 {\it Phys. Rev.} \textbf{177} 1857 and 1882; Agarwal G S and Wolf E 1970 {\it Phys. Rev. D} \textbf{2} 2161

\bibitem{Woo} Wootters W K 1987 {\it Ann. Phys. (N.Y.)} \textbf{176} 1 and  
2003 Picturing Qubits in Phase Space {\it Preprint} quant-ph/0306135; Feynman R P 1987 Negative Probabilities in {\it Quantum Implications. Essays
in Honour of David Bohm} Hiley B and Peat D  Eds. (Routledge, London)

\bibitem{MA} Mukunda N, Arvind, Chaturvedi S and Simon R 2004 {\it J. Math. Phys.} \textbf{45} 114; Mukunda N, Chaturvedi S and Simon R 2004 {\it Phys. Lett. A} \textbf{321} 160; Mukunda N, Marmo G, Zampini A, Chaturvedi S and
Simon R 2005 {\it  J. Math. Phys.} \textbf{46} 012106

\bibitem{Dir2} See Ref.\cite{Dir} and Dirac P A M 1947 {\it The Principles of Quantum Mechanics} (Oxford, at the Clarendon Press, 3d edition) Sect.32

\bibitem{Bar} Bargmann V 1964 {\it J. Math. Phys.} \textbf{5} 862

\bibitem{Ber} Berry M V 1984 {\it Proc. Roy. Soc. A (London)} \textbf{392} 45. The connection between Bargmann invariants and geometric phases is explored in: Mukunda N and Simon R 1993 {\it  Ann. Phys. (N.Y.)} \textbf{228} 205; Mukunda N, Arvind, Ercolessi E, Marmo G, Morandi G and Simon R 2003 {\it Phys. Rev. A} \textbf{67} 042114

\bibitem{FN1} The operators $\widehat{\mathcal{W}}\left(  q,p\right)
$ have been studied in: Mukunda N 1978 {\it  Pramana} \textbf{11} 1, where they have been called the elements of the ``Wigner Basis"  for the space of
operators. In Wootters W K (first of Refs. \cite{Woo}) they have
been called ``phase point operators".

\bibitem{Muk} Mukunda N in Ref. \cite{FN1}.

\end{thebibliography}
\end{document}